\documentclass[nofootinbib,showpacs,showkeys,preprintnumbers]{revtex4}%
\usepackage{amsfonts}
\usepackage{amsmath}
\usepackage{graphicx,color}
\usepackage{placeins}
\usepackage{footnote}
\usepackage{dcolumn}
\usepackage{soul}
\usepackage{ulem}
\usepackage{bm}
\usepackage{amssymb}
\usepackage{graphicx}%
\pdfoutput=1

\begin{document}
\title{Role of scalar dibaryon and $f_{0}(500)$ in the isovector channel of
low-energy neutron-proton scattering}
\author{Werner Deinet$^{a}$, Khaled Teilab$^{a,b}$, Francesco Giacosa$^{a,c}$, Dirk
H.\ Rischke$^{a}$}
\affiliation{$^{a}$~Goethe University Frankfurt am Main, Institute for Theoretical Physics,
Max-von-Laue-Str.\ 1, D-60438 Frankfurt am Main, Germany }
\affiliation{$^{b}$~Faculty of Science, Cairo University, 12613 Giza, Egypt }
\affiliation{$^{c}$~Institute of Physics, Jan Kochanowski University, ul.\ Swietokrzyska
15, 25-406 Kielce, Poland}

\begin{abstract}
We calculate the total and the differential cross section for $np$ scattering
at low energies in the isospin $I=1$ channel within the so-called extended
Linear Sigma Model. This model contains conventional (pseudo)scalar and
(axial--)vector mesons, as well as the nucleon and its chiral partner within
the mirror assignment. In order to obtain good agreement with {experimental data analysis
results} we need to consider two additional resonances: the lightest scalar
state $f_{0}(500)$ and a dibaryon state with quantum numbers $I=1,$
$J^{P}=0^{+}$ (a.k.a.\ $^{1}S_{0}$ resonance). The resonance $f_{0}(500)$ is
coupled to nucleons in a chirally invariant way through the mirror assignment
and is crucial for a qualitatively correct description of the shape of the
differential cross section. On the other hand, the dibaryon is exchanged in
the $s$--channel and is responsible of the large cross section close to
threshold. We compare our results to { data analysis results performed} by the SAID
program of the CNS Data Analysis Center {(in the following
"SAID results")}.

\end{abstract}

\maketitle

\normalem

\section{Introduction}

Nucleon-nucleon scattering at low energies has been investigated using
different effective approaches, see
e.g.\ Refs.\ \cite{machleidt,kaplan0,kaplan1,kaplan2,chen,kong,gegelia,epelbaum0,epelbaum1,vankolck,bedaquerev,coraggio,entem}
which are constructed according to the principles of chiral perturbation
theory (ChPT) \cite{chpt} [see also Ref.\ \cite{weisebook} and
refs.\ therein], the low-energy effective theory of the theory of strong
interactions, quantum chromodynamics (QCD). In ChPT, the chiral symmetry of
QCD is nonlinearly realized.

As initiated long ago in Ref.\ \cite{lee} [see also Ref.\ \cite{koch}],
another possibility to describe low-energy hadronic physics is based on the
linear realization of chiral symmetry via so-called Linear Sigma Models, see
e.g.\ Ref.\ \cite{geffen}. A modern variant is the extended Linear Sigma Model
(eLSM), which contains (pseudo)scalar and (axial--)vector mesons and which was
successfully applied in the context of meson-meson interactions
\cite{denisold,staniold,dick,staninew} and also meson-nucleon interactions
\cite{susannaold,susannanew,lisa}. In particular, baryons and their chiral
partners are treated in the so-called mirror assignment, in which a chirally
invariant mass term is present \cite{detar,zische,jido}. This is important,
since the smallness of the $\pi N$ sigma term implies that chiral symmetry
breaking alone cannot be responsible for the nucleon mass and other sources
(such as a gluon condensate) must exist which contribute to generating the
mass of the nucleon.

In this work, we use the eLSM in order to study nucleon-nucleon scattering. In
particular, we investigate neutron-proton scattering in the $I=1$ channel up
to a nucleon momentum of about $0.4$ GeV in the center of momentum (c.m.)
frame. In order to describe {experimental data analysis
results}, we need, apart from the usual quark-antiquark fields [see
e.g.\ Ref.\ \cite{isgur}], to also incorporate the light $f_{0}(500)$ meson
[for studies of this resonance see e.g.\ Refs.\ \cite{lowscalars,fariborz} as
well as the recent review \cite{sigmareview}]. As first shown in
Ref.\ \cite{susannaold} and then further investigated in
Refs.\ \cite{susagiu,achim2}, the resonance $f_{0}(500)$ can be coupled to the
eLSM in a chirally invariant way; the condensation of the field associated
with this resonance is then responsible for the emergence of the chirally
invariant mass term mentioned above. As shown in Ref.\ \cite{susagiu} by
studying nuclear matter saturation and in Ref.\ \cite{machleidtrev} by
studying the binding energy of nuclei, this resonance generates an attraction
between nucleons. In the present work we will confirm that its coupling to
nucleons is necessary for a reasonable description of neutron-proton
scattering data.

However, the exchange of mesons alone (even after the inclusion of
$f_{0}(500)$) is \textit{not} capable of describing the enhanced interaction
close to the neutron-proton threshold. As discussed previously in
Ref.\ \cite{kaplan0}, an additional resonance with baryon number $2$, isospin
$1$, as well as $J^{P}=0^{+}$ (equivalent to $^{1}S_{0}$ in the old
spectroscopic notation) {can be introduced to effectively describe
neutron-proton scattering}. Namely, this dibaryon resonance {[sometimes
called `dimeron' \cite{vankolck}]} is exchanged in the $s$-channel and
enhances considerably the cross section at threshold (up to a nucleon
c.m.\ momentum $p$ of about $0.2$ GeV). We will also determine the parameters
of this resonance, such as the nominal mass $m_{R}$ and width {and,
more importantly, we will investigate the existence of a pole in the complex
}$\sqrt{s}${ plane and estimate its position}. It turns out that the
on-shell tree-level decay width is larger than the difference $D_{R}\equiv
m_{R}-m_{p}-m_{n}$ of its mass from the threshold. As a consequence, this
state is not a conventional Breit-Wigner resonance due to strong threshold effects.

Indeed, in some of the previous works \cite{epelbaum0,epelbaum1,bedaquerev} the
dibaryon field was regarded as an auxiliary field that can be integrated out
in order to obtain an effective Lagrangian which contains only nucleonic
degrees of freedom. For the purpose of nucleon-nucleon scattering
phenomenology, this is certainly a reasonable and well-defined approach.
However, we believe that it is interesting to treat this state as a dibaryon
resonance. Moreover, as a consequence, we also expect an analogous resonance
in the neutron-neutron channel and possibly also in the proton-proton channel.

This paper is organized as follows: in Sec.\ II we present the model with
special attention to the resonance $f_{0}(500)$ and the dibaryon resonance. In
Sec.\ III we discuss our results for the $I=1$ neutron-proton total and
differential cross sections, step-by-step including various contributions. The
cross sections are compared to experimental data {analysis results}
from the SAID program of the CNS Data Analysis Center\footnote{Although SAID provides consistent
total and differential $np$ scattering cross sections summed over both
isospin channels, the individual $I=0$ and
$I=1$ differential $np$ scattering cross sections seem to be wrong by a factor
of two. In our analysis, we have taken this factor into account, i.e., we
divided SAID data by a factor of two.} \cite{SAID}.
Finally, in Sec.\ IV we give our conclusions and an outlook.

\section{The model}

The Lagrangian of the model used in our calculations has three parts:

\begin{enumerate}
\item[(i)] The mesonic part of the eLSM Lagrangian. This has been developed
and investigated for the two-flavor case $(N_{f}=2)$ in
Refs.\ \cite{denisold,staniold}, for the three-flavor case $(N_{f}=3)$ in
Refs.\ \cite{dick,staninew}, and recently for the four-flavor case $(N_{f}=4)$
in Ref.\ \cite{walaa}. For the explicit form of the Lagrangian, see the
aforementioned references.

\item[(ii)] The nucleonic part of the eLSM Lagrangian. For $N_{f}=2$, it
includes the interaction of the nucleon and its chiral partner $N^{\ast}$
(both states referred to as "nucleons" in the following) with $\bar{q}q$
mesons and a scalar isoscalar meson $f_{0}(500)$
\cite{susannaold,susagiu,susannanew} in a chirally invariant framework
[recently, the baryonic Lagrangian has been extended to $N_{f}=3$ in
Ref.\ \cite{lisa}]. In Sec.\ \ref{secelsm} we present the Lagrangian for
$N_{f}=2$ together with its parameters, while in Sec.\ \ref{secnpdiglag} we
consider only those terms which enter the calculation of nucleon-nucleon
scattering. Moreover, we also show how to include a form factor which
suppresses the interaction strength at high momenta.

\item[(iii)] The Lagrangian describing the interactions of two nucleons with
the $^{1}S_{0}$ dibaryon. This is constructed in Sec.\ \ref{sec1s0lag}.
\end{enumerate}

\subsection{The eLSM Lagrangian for nucleons}

\label{secelsm}

In the mirror assignment and in the two-flavor case, the eLSM Lagrangian in
the nucleon sector has the form \cite{susannaold}:
\begin{align}
\mathcal{L}_{eLSM}  &  =\bar{\Psi}_{1L}i\gamma_{\mu}{D}_{1L}^{\mu}\Psi
_{1L}+\bar{\Psi}_{1R}i\gamma_{\mu}{D}_{1R}^{\mu}\Psi_{1R}+\bar{\Psi}%
_{2L}i\gamma_{\mu}{D}_{2R}^{\mu}\Psi_{2L}+\bar{\Psi}_{2R}i\gamma_{\mu}{D}%
_{2L}^{\mu}\Psi_{2R}\nonumber\\
&  -\hat{g}_{1}(\bar{\Psi}_{1L}\Phi\Psi_{1R}+\bar{\Psi}_{1R}\Phi^{\dagger}%
\Psi_{1L})-\hat{g}_{2}(\bar{\Psi}_{2L}\Phi^{\dagger}\Psi_{2R}+\bar{\Psi}%
_{2R}\Phi\Psi_{2L})\nonumber\\
&  -a\chi(\bar{\Psi}_{1L}\Psi_{2R}-\bar{\Psi}_{1R}\Psi_{2L}-\bar{\Psi}%
_{2L}\Psi_{1R}+\bar{\Psi}_{2R}\Psi_{1L})\; , \label{eqL1}%
\end{align}
where:

\begin{enumerate}
\item[(i)] the first line of Eq.\ (\ref{eqL1}) describes the interaction of
the nucleons with (axial--)vector mesons via the derivatives $D_{1(2)L(R)}$,
which are defined as:
\begin{equation}
D_{1R}^{\mu}=\partial^{\mu}-ic_{1}R^{\mu}\text{ , }D_{1L}^{\mu}=\partial^{\mu
}-ic_{1}L^{\mu}\text{ },
\end{equation}%
\begin{equation}
D_{2R}^{\mu}=\partial^{\mu}-ic_{2}R^{\mu},\text{ }D_{2L}^{\mu}=\partial^{\mu
}-ic_{2}L^{\mu}\text{ }.
\end{equation}
The left-handed and right-handed fields $L^{\mu}$ and $R^{\mu}$ contain the
vector mesons ${{\omega}_{N}^{\mu}}$ and $\vec{\rho}^{\mu}$ and the
axial--vector mesons ${{f}_{1,N}^{\mu}}$ and $\vec{a}_{1}^{\mu}$:
\begin{equation}
L^{\mu}=({{\omega}_{N}^{\mu}}+{{f}_{1,N}^{\mu}}%
)t^{0}+(\vec{\rho}^{\mu}+\vec{a}_{1}^{\mu}) \cdot\vec{t}\text{ },
\end{equation}%
\begin{equation}
R^{\mu}=({{\omega}_{N}^{\mu}}-{{f}_{1,N}^{\mu}}%
)t^{0}+(\vec{\rho}^{\mu}-\vec{a}_{1}^{\mu})\cdot\vec{t}\text{ },
\end{equation}
where {$t^{0}$ and $\vec{t}$ represent the isospin matrices (}%
$t^{0}=1_{2}/2$ is half the $(2 \times2)$ unit matrix and $\vec{t}=\vec
{\sigma}/2,$ $\sigma_{i}$ being the $i$th Pauli matrix{)}. The
correspondence of the fields to quark-antiquark mesons listed in the PDG
\cite{pdg} is reported in Table I. Vector mesons are an important ingredient
for a good description of low-energy nucleon vacuum phenomenology, see
e.g.\ Refs.\ \cite{susannaold,dmitrasinovic}.

\item[(ii)] The second line of Eq.\ (\ref{eqL1}) describes the interaction of
the nucleons with the (pseudo)scalar mesons, parametrized in terms of the
matrix
\begin{equation}
\Phi=({{\sigma}_{N}}+i\eta_{N})t^{0}+(\vec{a}_{0}+i\vec{\pi}%
)\cdot\vec{t}\text{ ,}%
\end{equation}
see again Table I for the field-resonance correspondence. [Note that the field
$\eta_{N}$ has quark content $\left(  u\bar{u}+d\bar{d}\right)  /\sqrt{2}$ and
can be expressed as a combination of the physical fields $\eta$ and
$\eta^{\prime}$ as: $\eta_{N}=\cos\varphi_{P}\eta-\sin\varphi_{P}\eta^{\prime
}$ where the mixing angle is $\varphi_{P}\approx-44^{\circ}$ \cite{dick}; in
the other sectors we neglect the small strange-nonstrange mixing.] The
interaction terms in the second line provide a contribution to the nucleon
masses via the condensation of $\sigma$ ($\sigma\rightarrow\sigma+\phi,$ where
$\phi$ is the chiral condensate). In the original Linear Sigma Model, this was
the only contribution to the nucleon mass in the chiral limit. {Note
that the resonances }$a_{0}(980),$ $f_{0}(980),${ and }%
$K_{0}^{\ast}(800)${ are not included in the model since they turn out
to be predominantly four-quark objects, see e.g.\ Refs.\
\cite{lowscalars,giacosatqmix,thomas,milena} and refs.\ therein. Namely, these
resonances form, together with the resonance }$f_{0}(500),$ { a nonet of
non-conventional mesons. As we discuss below, in the }$N_{f}=2$ {
framework adopted in this work, only }$f_{0}(500)$ { shall be considered
[since it has no (open or hidden) strangeness content in its wave function].}

\item[(iii)] The third and last line of Eq.\ (\ref{eqL1}) describes the
interaction of the nucleon fields $\Psi_{1}$ and $\Psi_{2}$ with the scalar
non-conventional meson $\chi$. The latter gives a contribution to the nucleon
masses due to the condensation of $\chi$ ($\chi\rightarrow\chi+\chi_{0}$). The
mass parameter
\begin{equation}
m_{0}=a\chi_{0} \label{eqm0}%
\end{equation}
was discussed in the pioneering work of Ref.\ \cite{detar} and further
investigated in Refs.\ \cite{zische,jido,susannaold,susannanew,lisa}. In
Ref.\ \cite{susagiu} it was suggested that the mass term $m_{0}$ arises from
the condensation of the scalar isoscalar field $\chi$. The latter corresponds
to the resonance $f_{0}(500)$ in the context of nuclear physics
\cite{susagiu,achim2}.
\end{enumerate}

\begin{center}
\textbf{Table I}: Correspondence of the fields to mesons listed in
Ref.\ \cite{pdg}.%

\begin{tabular}
[c]{|c|c|c|c|c|c|}\hline
Field & PDG & Quark content & $I$ & $J^{PC}$ & Mass (GeV)\\\hline
$\pi^{+},\pi^{-},\pi^{0}$ & $\pi$ & $u\bar{d},d\bar{u},\frac{u\bar{u}-d\bar
{d}}{\sqrt{2}}$ & $1$ & $0^{-+}$ & $0.13957$\\\hline
$\eta$ & $\eta(547)$ & $\frac{u\bar{u}+d\bar{d}}{\sqrt{2}}\cos\varphi
_{P}-s\bar{s}\sin\varphi_{P}$ & $0$ & $0^{-+}$ & $0.54786$\\\hline
$\eta^{\prime}$ & $\eta^{\prime}(958)$ & $\frac{u\bar{u}+d\bar{d}}{\sqrt{2}%
}\sin\varphi_{P}+s\bar{s}\cos\varphi_{P}$ & $0$ & $0^{-+}$ & $0.95778$\\\hline
$a_{0}^{+},a_{0}^{-},a_{0}^{0}$ & $a_{0}(1450)$ & $u\bar{d},d\bar{u}%
,\frac{u\bar{u}-d\bar{d}}{\sqrt{2}}$ & $1$ & $0^{++}$ & $1.474$\\\hline
$\sigma_{N}$ & $f_{0}(1370)$ & $\frac{u\bar{u}+d\bar{d}}{\sqrt{2}}$ & $0$ &
$0^{++}$ & $1.350$\\\hline
$\rho^{+},\rho^{-},\rho^{0}$ & $\rho(770)$ & $u\bar{d},d\bar{u},\frac{u\bar
{u}-d\bar{d}}{\sqrt{2}}$ & $1$ & $1^{--}$ & $0.77526$\\\hline
$\omega_{N}$ & $\omega(782)$ & $\frac{u\bar{u}+d\bar{d}}{\sqrt{2}}$ & $0$ &
$1^{--}$ & $0.78265$\\\hline
$a_{1}^{+},a_{1}^{-},a_{1}^{0}$ & $a_{1}(1230)$ & $u\bar{d},d\bar{u}%
,\frac{u\bar{u}-d\bar{d}}{\sqrt{2}}$ & $1$ & $1^{++}$ & $1.230$\\\hline
$f_{1,N}$ & $f_{1}(1285)$ & $\frac{u\bar{u}+d\bar{d}}{\sqrt{2}}$ & $0$ &
$1^{++}$ & $1.2819$\\\hline
$\chi$ & $f_{0}(500)$ & $\pi\pi$ or $[u,d][\bar{u},\bar{d}]$ & $0$ & $0^{++}$ &
$0.475$\\\hline
\end{tabular}

\end{center}

Finally, the nucleon fields $\Psi_{1}$ and $\Psi_{2}$ are related to the
physical states of the nucleon $N$ and its chiral partner $N^{\ast}$ as:
\begin{equation}
\Psi_{1}=\frac{1}{\sqrt{2\cosh\delta}}\left(  Ne^{\delta/2}+\gamma_{5}N^{\ast
}e^{-\delta/2}\right)  \text{ },
\end{equation}%
\begin{equation}
\Psi_{2}=\frac{1}{\sqrt{2\cosh\delta}}\left(  \gamma_{5}Ne^{-\delta/2}%
-N^{\ast}e^{\delta/2}\right)  \text{ },
\end{equation}
where
\begin{equation}
\cosh\delta=\frac{m_{N}+m_{N^{\ast}}}{2m_{0}}\text{ }.
\end{equation}
The field $N$ corresponds to the nucleon $N(939)$ while $N^{\ast}$ to its
chiral partner, which could be $N(1535)$ or $N(1650)$. For the purposes of the
present work, the assignment of $N^{\ast}$ is not crucial. For the sake of
definiteness, we will use the results of Ref.\ \cite{susannanew}, in which
$N(1650)$ is regarded as the chiral partner. On the other hand, in an enlarged
mixing scenario \cite{lisa}, $N(1535)$ is favored as the chiral partner of the
nucleon. However, using this alternative scenario does not lead to noticeable
quantitative changes of our results.

The masses of the nucleon $N$ and its chiral partner $N^{\ast}$ are given by:
\begin{equation}
m_{N,N^{\ast}}=\sqrt{m_{0}^{2}+\frac{(\hat{g_{1}}+\hat{g_{2}})^{2}}{16}
\phi^{2} } \pm\frac{1}{4}(\hat{g_{1}}-\hat{g_{2}})\phi\text{ }.
\end{equation}
In the limit $m_{0} \rightarrow0$, one obtains the result $m_{N}=\hat{g_{1}%
}\phi/2$, i.e., the nucleon mass is solely generated by the chiral condensate
[as in the original Linear Sigma Model \cite{lee,koch}].

Using the Lagrangian of Eq.\ (\ref{eqL1}) we also obtain expressions for the
axial coupling constants $g_{A}^{N}$ and $g_{A}^{N^{\ast}}$ of the nucleon and
its chiral partner $N^{\ast}$, respectively,
\begin{equation}
g_{A}^{N}=\frac{1}{2\cosh\delta}\left(  g_{A}^{(1)}e^{\delta}+g_{A}%
^{(2)}e^{-\delta}\right)  \text{ },\quad\quad g_{A}^{N^{\ast}}=\frac{1}%
{2\cosh\delta}\left(  g_{A}^{(1)}e^{-\delta}+g_{A}^{(2)}e^{\delta}\right)
\text{ },
\end{equation}
where
\begin{equation}
g_{A}^{(1)}=1-\frac{c_{1}}{g_{1}}\left(  1-\frac{1}{Z^{2}}\right)  \text{
},\quad\quad g_{A}^{(2)}=-1+\frac{c_{2}}{g_{1}}\left(  1-\frac{1}{Z^{2}%
}\right)  \text{ .}%
\end{equation}
We recall that $Z=(1-g_{1}w\phi)^{-1/2}=1.67>1$, where $g_{1}=5.84$ describes
the coupling constant of (pseudo)scalar and (axial--)vector mesons, and
$w=g_{1}\phi/m_{a_{1}}^{2}$. This parameter arises from the mixing of
pseudoscalar and axial--vector mesons, see Refs.\ \cite{denisold,dick}. As a
consequence, the condensate reads $\phi=Zf_{\pi},$ where $f_{\pi}=0.0924$ GeV
is the pion decay constant. The importance of vector mesons is evident, since
only for nonzero $c_{1}$ and $c_{2}$ (and for $Z>1$), it is possible to get an
agreement of the axial coupling constants with experimental data and
lattice-QCD calculations \cite{takahashi}.

In total, the nucleon part of the model has five independent parameters
($a,\hat{g}_{1},\hat{g}_{2},c_{1},c_{2}$), which are determined by using the
PDG values $m_{N}=0.939$ GeV, $m_{N^{\ast}}=1.650$ GeV, $\Gamma_{N^{\ast
}\rightarrow NP}=0.128$ GeV$,$ the axial coupling constant $g_{A}^{N}=1.267,$
as well as lattice-QCD calculations of the axial coupling constant
$g_{A}^{N^{\ast}}=0.55$ \cite{takahashi}, for details and determination of the
errors, see Ref.\ \cite{susannaold}. Explicitly:
\begin{equation}
c_{1}=-3.34\text{ },\text{ }c_{2}=14.74\text{ },\text{ }\hat{g}_{1}=9.47\text{
}, \text{ }\hat{g}_{2}=18.69\text{ },\text{ }m_{0}=0.704\text{ GeV .}%
\end{equation}

Finally, as described in Ref.\ \cite{susagiu}, the condensate $\chi_{0}$ takes
the form $\chi_{0}=g_{\chi\pi\pi}\phi^{2}/m_{\chi}^{2},$ where $g_{\chi\pi\pi
}$ is the $\chi\pi\pi$ coupling constant \cite{giacosatqmix,achim1}. Its
numerical value was determined to be $0.45$ GeV \cite{susagiu} by requiring a
correct description of the nuclear matter ground state. Since we assign
$\chi\equiv f_{0}(500)$, we use $m_{\chi}=(0.475\pm0.25)$ GeV \cite{pdg}. As a
consequence of $\chi_{0}=g_{\chi\pi\pi}\phi^{2}/m_{\chi}^{2}$, the constant
$a$ in Eq.\ (\ref{eqL1}) reads:
\begin{equation}
a=\frac{m_{0}}{\chi_{0}}=\frac{m_{0}m_{\chi}^{2}}{g_{\chi\pi\pi}(Zf_{\pi}%
)^{2}}\text{ }, \label{eqfora}%
\end{equation}
{which equals }${14.8}${ for $m_{\chi}=$}${0.475}%
${~GeV.}

The value of $a$ as given by Eq.\ (\ref{eqfora}) is the maximum value for the
coupling of $\chi$ to nucleons. If other scalar condensates (e.g.\ a glueball
condensate) contribute to the mass parameter $m_{0}$, Eq.\ (\ref{eqm0}), the
value of $a$ would be reduced. Hence, for the results presented in
Sec.\ \ref{res} we will choose also lower values for $a$ than given by
Eq.\ (\ref{eqfora}), if necessary to achieve good agreement with { SAID results}.

\subsection{Lagrangian for nucleon--nucleon elastic scattering}

\label{secnpdiglag}

\begin{figure}[ptb]
\centering
\includegraphics[width=5cm]{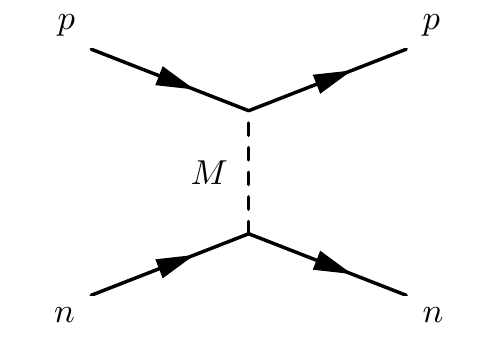} \qquad
\includegraphics[width=5cm]{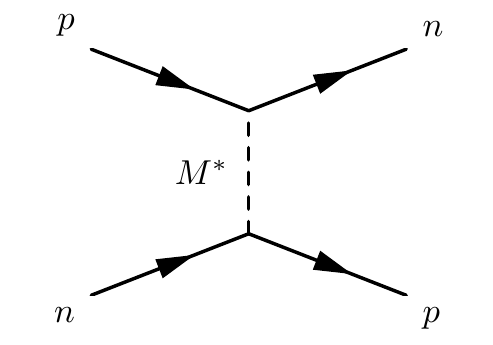} \caption{Feynman diagrams for
$np$ scattering in the eLSM. Left: neutral meson exchange, $M\equiv\chi
,\sigma,a^{0}_{0},\pi^{0},\eta,\omega,\rho^{0},f_{1},a^{0}_{1}$. Right:
charged meson exchange, $M^{*}\equiv a^{\pm}_{0},\pi^{\pm},\rho^{\pm},a^{\pm
}_{1}$.}%
\label{figfeyn1}%
\end{figure}

Only some of the terms contained in Eq.\ (\ref{eqL1}) contribute to elastic
nucleon-nucleon scattering (for instance, the nucleon resonance $N^{\ast}$
does not contribute). We thus split the full Lagrangian $\mathcal{L}_{eLSM}
=\mathcal{L}_{NN}+\mathcal{L}_{rest},$ where the relevant terms for our
calculations are contained in $\mathcal{L}_{NN}$. Its explicit form in terms
of physical fields reads:
\begin{align}
\mathcal{L}_{NN}  &  =\frac{1}{2\cosh\delta}\big(e^{\delta}c_{1}\overline
{N}\left\{  {{\omega}_{N}^{\mu}}t^{0}+\vec{\rho}^{\mu}\cdot\vec{t}
+\left[  {{f}_{1,N}^{\mu}}t^{0} +\vec{a}_{1}^{\mu}\cdot\vec
{t}+wZ\left(  \partial_{\mu}\eta_{N}t^{0} +\partial_{\mu}\vec{\pi}\cdot\vec
{t}\; \right)  \right]  \gamma_{5}\right\}  \gamma_{\mu}N\nonumber\\
&  \hspace*{1.4cm} +e^{-\delta}c_{2}\overline{N}\left\{  {{\omega
}_{N}^{\mu}}t^{0} +\vec{\rho}^{\mu}\cdot\vec{t} -\left[  {{f}%
_{1,N}^{\mu}}t^{0}+\vec{a}_{1}^{\mu}\cdot\vec{t}+wZ\left(  \partial_{\mu}%
\eta_{N}t^{0}+\partial_{\mu}\vec{\pi}\cdot\vec{t} \; \right)  \right]
\gamma_{5}\right\}  \gamma_{\mu}N\nonumber\\
&  \hspace*{1.4cm}-e^{\delta}\hat{g}_{1}\overline{N}\left\{  \left[  \left(
{{\sigma}_{N}} +\varphi\right)  t^{0}+\vec{a}_{0}\cdot\vec{t}\;
\right]  +i Z\left(  \eta_{N}t^{0}+\vec{\pi}\cdot\vec{t}\; \right)  \gamma
_{5}\right\}  N\nonumber\\
&  \hspace*{1.4cm}+e^{-\delta}\hat{g}_{2}\overline{N}\left\{  \left[  \left(
{{\sigma}_{N}} +\varphi\right)  t^{0}+\vec{a}_{0}\cdot\vec{t}\;
\right]  -iZ\left(  \eta_{N}t^{0}+\vec{\pi}\cdot\vec{t}\; \right)  \gamma
_{5}\right\}  N\nonumber\\
&  \hspace*{1.4cm} -2a\overline{N}\left(  \chi+\chi_{0}\right)  N\big) \;.
\label{eqLNN}%
\end{align}
The resulting $t$-- and $u$--channel Feynman diagrams for nucleon-nucleon
interactions via meson exchange are shown in Fig.\ \ref{figfeyn1}. We use the
following propagators for the exchanged mesons:
\begin{equation}
G_{S}=\frac{i}{q^{2}-m_{i}^{2}}\;, \qquad G_{V,\alpha\beta}=-i\left(
g_{\alpha\beta}-\frac{q_{\alpha}q_{\beta}}{m_{i}^{2}}\right)  \frac{1}%
{q^{2}-m_{i}^{2}}%
\end{equation}
for spinless and spin-$1$ particles, respectively; $m_{i}$ denotes the
on-shell mass of the exchanged meson.

As a last point, we describe the introduction of form factors. The model
described in Eq.\ (\ref{eqL1}) is not a fundamental model which describes the
interactions of point-like particles, but an effective model whose degrees of
freedom are hadrons (nucleons and mesons) which have a finite extension
($\sim0.5$ fm). Therefore, the tree-level diagrams derived from the Lagrangian
(\ref{eqL1}) are valid when the momentum exchanged at a certain vertex is
smaller than $\sim2\;\mathrm{fm}^{-1}\simeq0.4$ GeV. Therefore, as various
works have shown, see e.g.\ Refs.\ \cite{close,dong,kaptari,wang} and also
Ref.\ \cite[see section "Quark model in Standard Model and Related
Topics"]{pdg}, and as we shall also see later on, it is important to introduce
a form factor which reduces the interaction strength when the momenta of the
hadrons are large. In this work we will use the following form factor attached
to each nucleon-nucleon-meson vertex:
\begin{equation}
F(q^{2})=\exp\left(  -\frac{|q^{2}-m_{i}^{2}|}{\Lambda_{cut}^{2}}\right)
\text{ }, \label{ff}%
\end{equation}
where $q^{2}$ is the {square of the four-}momentum transfer involved in
the process ($q^{\mu}$ is the four-momentum of the exchanged meson, and
$m_{i}$ its mass, $i=\pi,\rho,\ldots$). The parameter $\Lambda_{cut}$ is a
hadronic energy scale, which is $\sim1$ GeV.

{An alternative approach to form factors is the implementation of
unitarization approaches (such as the so-called K-matrix unitarization) which
also cause a decrease of cross section at high energies. However, their use
would imply the need for a partial-wave analysis which goes beyond the scope of
the present paper. We leave this study as well as the analysis of
neutron-proton scattering in all partial waves [see e.g.\ Ref.\ \cite{coraggio}]
for the future.}

\subsection{Interaction Lagrangian for the $^{1}S_{0}$ dibaryon}

\label{sec1s0lag}

The interaction of nucleons via meson exchange is not capable of describing
the large cross section close to threshold. Namely, the interaction strength
is three orders of magnitude larger than what can be achieved through meson
exchange: a neutron-proton resonance is responsible for the enhanced cross section.

In order to describe this resonance within our framework, we introduce a new
field, denoted as $\Phi_{R}$, which has quantum numbers $I=1$, $J^{P}=0^{+}$
(a.k.a.\ $^{1}S_{0}$) and contributes to $I=1$ $np$ scattering close to
threshold. The wave function of the $I_{z}=0$ component (of relevance for the
following) is given by
\begin{equation}
\left\vert \Phi_{R}\right\rangle =\left\vert \text{space: ground state}
\right\rangle \left\vert \uparrow\downarrow-\downarrow\uparrow\right\rangle
\left\vert np+pn\right\rangle \text{ .}%
\end{equation}
Being part of an isospin multiplet, there are two analogous states,
$\left\vert pp\right\rangle $ with $I_{z}=1$, and $\left\vert nn\right\rangle
$ with $I_{z}=-1$, see the corresponding discussion in Sec.\ IV.

\begin{figure}[ptb]
\centering
\includegraphics[width=5cm]{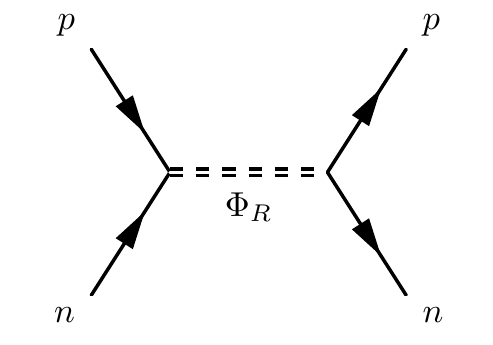}\caption{Feynman diagram for $I=1$
$np$ scattering via the {charged }$^{1}S_{0}$ resonance.}%
\label{fig:schannel}%
\end{figure}

The Lagrangian coupling $\Phi_{R}$ to nucleons is given by:
\begin{equation}
\mathcal{L}_{R}=iG_{R}\left(  N^{T}C\gamma^{5}\Phi_{R}{t^{1}}N+\bar{N}%
\gamma^{5} {\Phi}_{R}^{\ast}{t^{1}}C\bar{N}^{T}\right)  \text{ ,} \label{eqLR}%
\end{equation}
where $C$ denotes the charge-conjugation matrix. The first term on the
right-hand side of Eq.\ (\ref{eqLR}) describes the two incoming nucleons
creating the dibaryon, while the other term describes the decay of the
dibaryon into two outgoing nucleons. A similar Lagrangian for the deuteron was
studied in Ref.\ \cite{dong}. The corresponding Feynman diagram for $np$
scattering is shown in Fig.\ \ref{fig:schannel}. 

Using Eq.\ (\ref{eqLR}) we calculate the width $\Gamma(p)$ of the $^{1}S_{0}$
state as:
\begin{equation}
\Gamma(p)=\frac{G_{R}^{2}p}{4{\pi}}\text{ },
\end{equation}
where $p$ denotes the modulus of the three-momentum of an outgoing nucleon.
The propagator of the $^{1}S_{0}$ state is given by:
\begin{equation}
\Delta_{R}(s)=\frac{i}{s-m_{R}^{2}+i\sqrt{s}\,\Gamma(p)}\text{ ,}
\label{propr1}%
\end{equation}
where $m_{R}$ denotes the `mass' of the dibaryon resonance. We recall that $p$
is a function of the kinematic variable $s$:
\begin{equation}
p=p(s)=\sqrt{\frac{s^{2}+(m_{p}^{2}-m_{n}^{2})^{2}-2s(m_{p}^{2}+m_{n}^{2}%
)}{4s}}\text{ ,}%
\end{equation}
where $m_{p}$ and $m_{n}$ are the proton and the neutron masses, respectively.
For a good description of  {SAID results} it is essential to
consider the decay width as a function of $p$, i.e., $\Gamma(p)$. Setting the
decay width to a constant, $\Gamma_{R}\equiv\Gamma(p_{R})$, where $p_{R}\equiv
p(m_{R}^{2})$, (i.e., the Breit-Wigner limit) is definitely not a good
approximation in the present context. Hence, the quantity $m_{R}$ should not
be regarded as a conventional resonance mass, but as a parameter corresponding
to the root of the real part of the denominator of the propagator, see also
the discussion in Sec.\ \ref{scaldibonly}.

{At the end of this section, two comments are in order:}

{(i) The propagator (\ref{propr1}) emerges upon a resummation of
proton-neutron loops. In this respect, it corresponds to a (partial)
unitarization in the }$s$-{channel for this particular process. Note,
for simplicity the real part of the propagator's denominator has not been
modified (see Sec.\ III.A). }

{(ii) As discussed in Refs.\ \cite{machleidt,vankolck,coraggio}, it is
not necessary to introduce an additional field }$\Phi_{R}$ {to describe
data. One would obtain an equally good description by starting with a quartic
interaction term proportional to }$\left(  N^{T}C\gamma^{5}{t^{1}}N\right)
^{2}${ and by doing a resummation of the proton-neutron loop emerging
from it. This }$S${-wave resummation generates an expression which
resembles that of a propagator of a scalar particle. Then, in the framework of
an correct description of data, the inclusion of an explicit dimeron field
}$\Phi_{R}\ ${is possible but not necessary. Yet, the point that we
would like to address is if a pole in the complex plane on the second Riemann
sheet exists. Namely, this is the condition that should be met for a state
to exist. In fact, the position of the pole is independent on the particular
process and (in principle) is also independent on the particular Lagrangian
employed, as long as data are correctly described. Indeed, we show in the next
section that we do find a pole in the complex plane. }

\section{Results}

\label{res}

We now turn to the results. We present them successively including more
ingredients: (i) we consider only the scalar dibaryon [Eq.\ (\ref{eqLR})];
(ii) we consider a reduced model with the dibaryon and the scalar meson
$\chi\equiv f_{0}(500)$ [Eq.\ (\ref{eqLR}) and the last line of
Eq.\ (\ref{eqLNN})]; (iii) we include all other mesons without form factor
[Eqs.\ (\ref{eqLNN}) and (\ref{eqLR}) with $\Lambda_{cut}\rightarrow\infty$];
(iv) we include also a form factor [Eqs.\ (\ref{eqLNN}) and (\ref{eqLR}) with
finite $\Lambda_{cut}$].

In all cases, the cross section for $np$ scattering in the $I=1$ channel was
calculated by splitting the scattering amplitude $\mathcal{M}$ into two parts,
$\mathcal{M}_{I=0}$ and $\mathcal{M}_{I=1}$, according to the formalism
presented in Ref.\ \cite{goldberger}. {The masses of the neutron and
proton were set equal to }${938.919}${~MeV (the average of
both masses), except for the study of the pole and spectral function of the
$^{1}S_{0}$ dibaryon resonance, which was done using the masses reported in Ref.\
\cite{pdg}}. The results were cross-checked by calculating the cross section
for $pp$ scattering {neglecting Coulomb interaction} using the programs
FeynRules 2.0 \cite{feynrules} and MadGraph 2.3.0 \cite{madgraph}. Results of
both calculations were identical within numerical precision.

\subsection{Scalar dibaryon only}

\label{scaldibonly}

We first consider the case where only the Lagrangian (\ref{eqLR}) is
considered. The interaction is mediated by the dibaryon resonance $\Phi_{R}$.
Fig.\ 3(a) shows the total cross section for different values
of $D_{R}=m_{R}-m_{p}-m_{n}$, where $m_{R}$ is the mass parameter entering
Eq.\ (\ref{propr1}). As one observes, the total cross section is very large at
threshold and drops rapidly with increasing momentum. The best agreement with
 { SAID results} is obtained for the dibaryon coupling strength
$G_{R}= 2.13$ and for $D_{R}=0.0015$ GeV. Namely, we obtain a good description
of the  { SAID results} over three orders of magnitudes up to a
c.m.\ momentum of about $0.2$~GeV. Clearly, the calculation using only the
$^{1}S_{0}$ resonance cannot describe the  { SAID results} at
higher momenta. For momenta above $0.2$~GeV, the interaction via meson
exchange (as, for instance, described via the eLSM Lagrangian) dominates.

As expected, there is no angular dependence (at any $p$) of the theoretically
calculated differential cross section when only the dibaryon is included, see
 Fig.\ 3(b). On the contrary, { SAID results} show
an enhancement at forward and backward angles. This enhancement increases with
momentum and, as we shall see, can be explained considering meson exchange (in
particular $f_{0}(500)$).

\begin{figure}[ptb]
\includegraphics[width=0.47\linewidth]{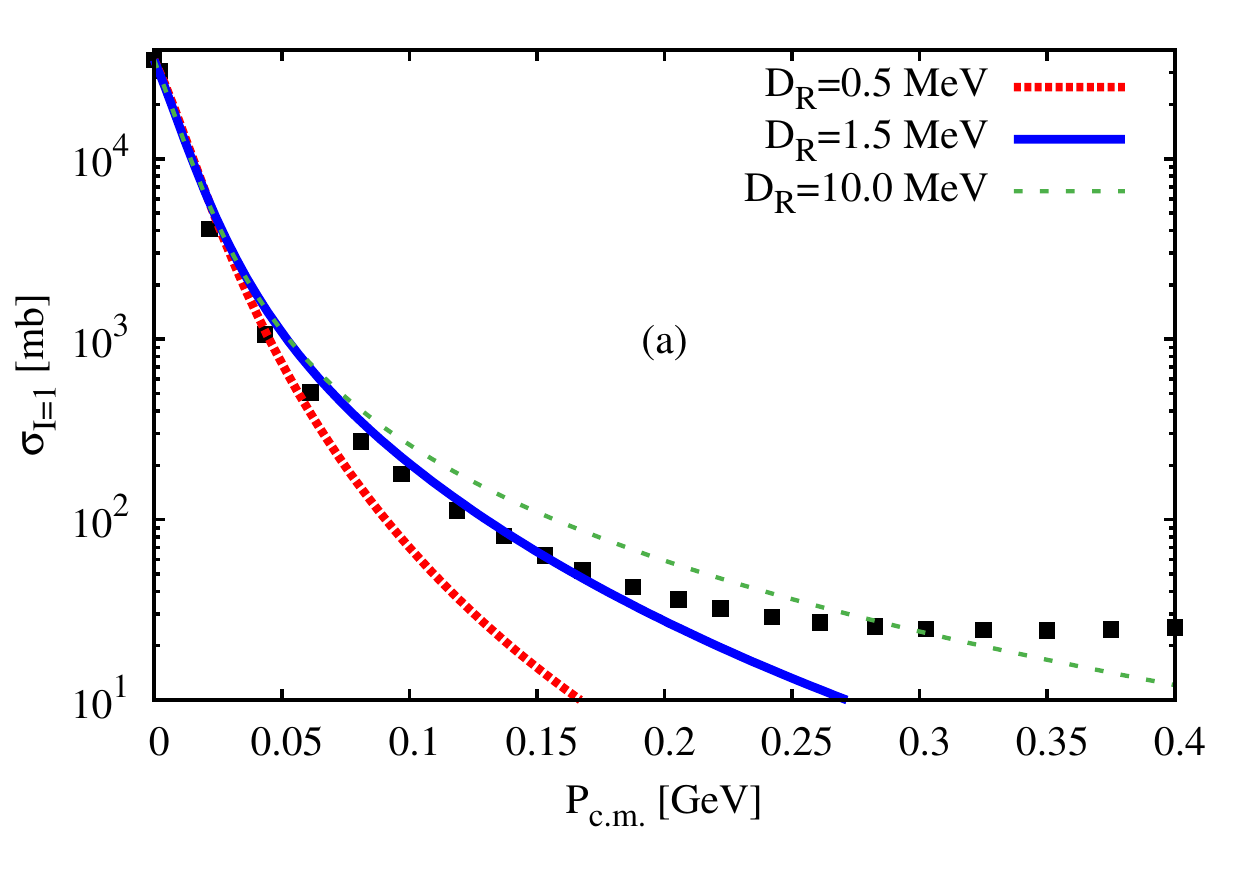} \quad
\includegraphics[width=0.47\linewidth]{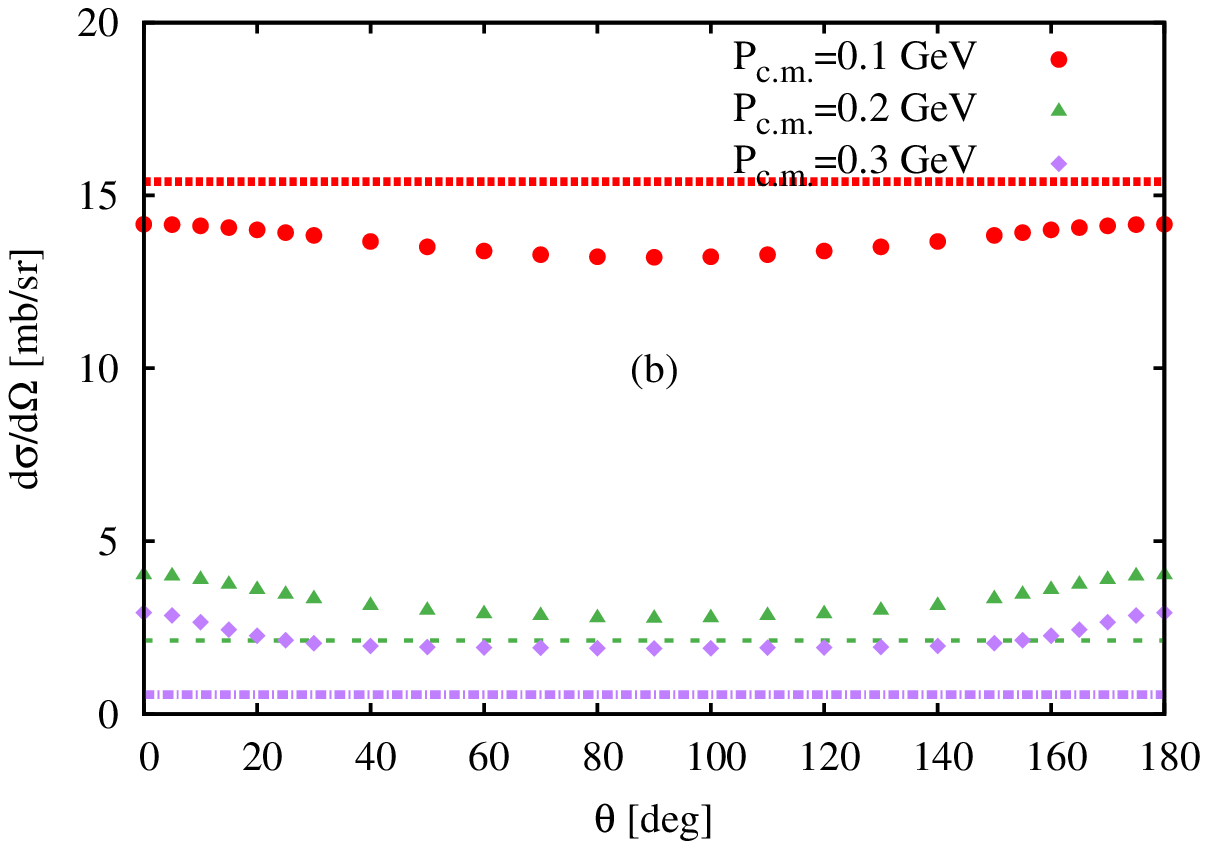}\caption{Total (a)
and differential (b) cross section for $I=1$ $np$ scattering via the
$^{1}S_{0}$ resonance only. In Fig.\ a, the total cross section is
shown for different values of $D_{R}= 0.0005$ GeV (red dotted curve), $0.0015$
GeV (blue solid curve), and $0.01$ GeV (green dashed curve). The corresponding
values for $G_{R}$ (chosen to fit the total cross section at threshold) are
$1.23$, $2.13$, and $5.5$, respectively. In Fig.\ b, the differential
cross section is shown for $D_{R}=0.0015$ GeV and $G_{R}=2.13$ for nucleon
c.m.\ momenta 0.1 GeV (red dotted curve), 0.2 GeV (green dashed curve), and
0.3 GeV (magenta dash-dotted curve), respectively. Data points are taken from
the SAID program \cite{SAID}.}%
\label{fig:Resonly}%
\end{figure}

In conclusion, we find that there is an isotriplet dibaryon resonance with
nominal mass $m_{R}=m_{p}+m_{n}+0.0015$ GeV. The corresponding on-shell decay
width is $\Gamma_{R}=0.0135$ GeV, which is much larger than $D_{R}$. Whenever
the tree-level decay width is comparable to or larger than the distance of the
mass from the threshold, we are not dealing with a standard resonance, see
e.g.\ Ref.\ \cite{lupo} and refs.\ therein. Here, the situation is even more
extreme, since $\Gamma_{R}\gg D_{R}$. This is also why many authors were
extremely careful in discussing this putative state as a standard resonance.

In the literature, it is common to investigate the existence and position of
pole(s) in the complex $\sqrt{s}$ plane, especially in presence of wide
resonances [as, for instance, in the renowned case of the resonance
$f_{0}(500),$ see Ref.\ \cite{sigmareview} and refs.\ therein]. To this end,
we investigate the presence of a pole by using the formalism discussed in
Ref.\ \cite{thomas}: we introduce a form factor in the decay width,
$\Gamma(p)\rightarrow\Gamma_{\Lambda_{R}}(p)=\Gamma(p)e^{-2p^{2}/\Lambda
_{R}^{2}}$. Then, according to the optical theorem, the one-loop self-energy
$\Sigma(s)$ (which consists of a neutron-proton loop) fulfills
$\operatorname{Im}\Sigma(s)=\sqrt{s}\Gamma_{\Lambda_{R}}(p)$. The real part is
determined by using the dispersion integral
\begin{equation}
\operatorname{Re}\Sigma(s)=-\frac{1}{\pi}\;\mathrm{P.V.}\int_{m_{p}+m_{n}%
}^{\infty}ds^{\prime}\,\frac{\operatorname{Im}\Sigma(s^{\prime})}{s-s^{\prime
}}\text{ ,}%
\end{equation}
where P.V.\ stands for principal value. The dressed propagator of the dibaryon
reads
\begin{equation}
\Delta_{R}^{\text{dressed}}(s)=\frac{1}{s-m_{R}^{2}+\operatorname{Re}%
\Sigma(s)-\operatorname{Re}\Sigma(m_{R}^{2})+i\operatorname{Im}\Sigma
(s)}\text{ ,} \label{propdressed}%
\end{equation}
while its spectral function is given by
\begin{equation}
d_{R}(\sqrt{s})=\frac{2\sqrt{s}}{\pi}\operatorname{Im}\Delta_{R}%
^{\text{dressed}}(s)\text{ .}%
\end{equation}
The latter is plotted in Fig.\ 4 using $\Lambda_{R}=0.5$ GeV [close to the
values obtained in Refs.\ \cite{thomas,milena}]. One notices a peak very close
to threshold (only $0.0000174$ GeV away from it) and then a rapid descent.
Note that the peak does not correspond to the nominal mass $m_{R}$. Since
$\Lambda_{R}$ is a free parameter, we also show the spectral function for
$\Lambda_{R}=0.3$ GeV. The quantitative difference to the previous case is
small, the qualitative features remain the same.

{Note that, in principle, one should have used from the very beginning
the propagator (\ref{propdressed}) in the calculation of the cross section.
Such a calculation would require to
determine the parameter }$G_{R}$ {in a set of complicated coupled equations. 
In view of the uncertainty on }$\Lambda_{R},$ {such a
procedure, while formally correct, goes beyond the scope of the present work.
However, we have a posteriori verified that the full propagators
(\ref{propr1}) and (\ref{propdressed}) deliver similar results for the cross
section.}

Finally, we turn to the position of the pole of the dimeron. For $\Lambda
_{R}=0.5$ GeV we find a pole for
\begin{equation}
\sqrt{s_{\text{pole}}}=m_{p}+m_{n}+0.014\text{ GeV}\text{ }-i\;0.0774\text{
GeV}\;.
\end{equation}
We observe that the decay width associated with the imaginary part of the pole
is much larger than the tree-level decay width: $\Gamma_{pole}=0.1548$ GeV. In
addition, the pole mass, being $0.014$ GeV above the threshold, is larger. For
this very peculiar resonance there is no correspondence between nominal mass,
peak of the spectral function, and pole mass.

\begin{figure}[ptb]
\begin{center}
\includegraphics[width=0.5\linewidth]{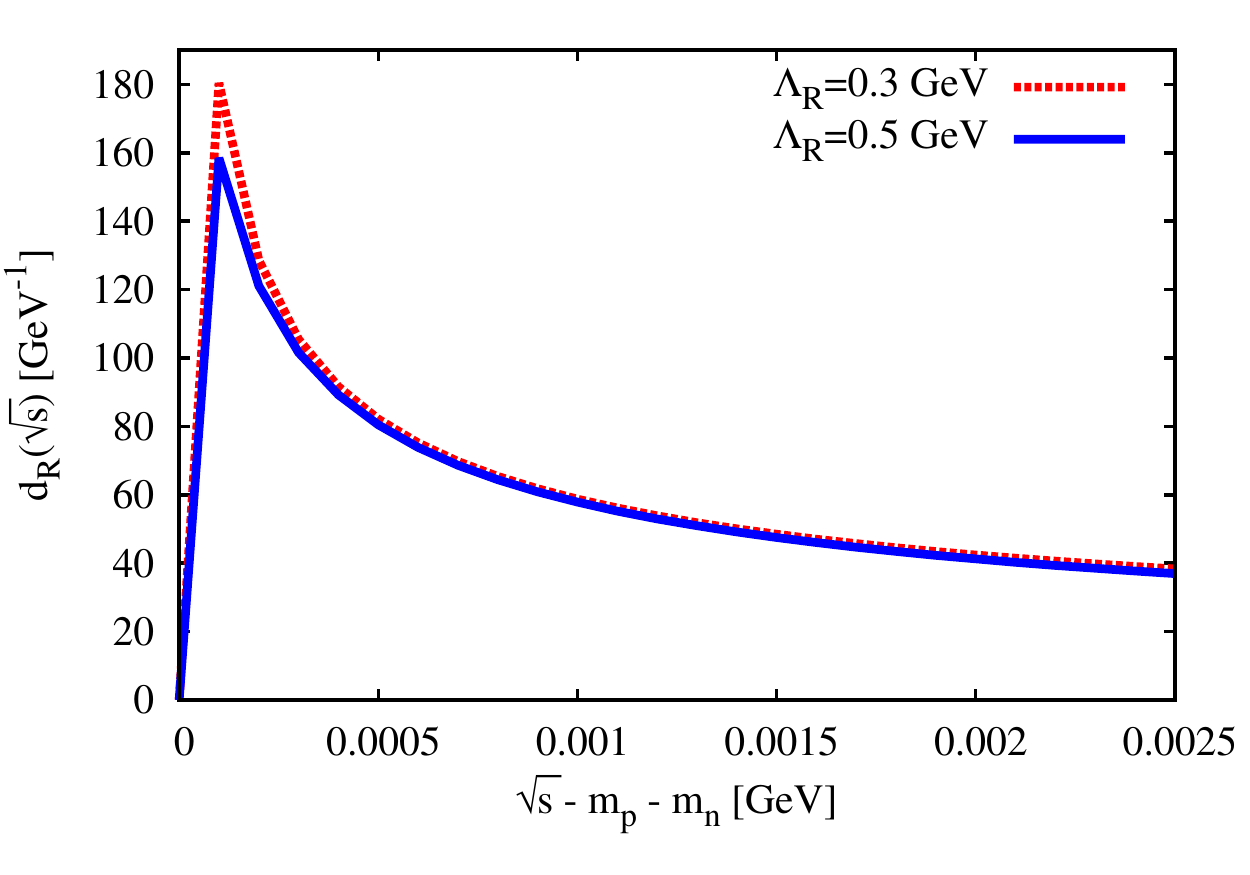}
\end{center}
\caption{Spectral function of the scalar dibaryon as function of $\sqrt
{s}-m_{p}-m_{n}$ for $\Lambda_{R}=0.5$ GeV and $\Lambda_{R}=0.3$ GeV. We
verified that the spectral functions are correctly normalized to unity.}%
\end{figure}

For $\Lambda_{R}=0.3$ GeV, the pole is located at
\begin{equation}
\sqrt{s_{\text{pole}}}=m_{p}+m_{n}+0.0273\text{ GeV}\text{ }-i\;0.0309\text{
GeV}\text{ ,}%
\end{equation}
which has a larger mass but a smaller width. While the spectral function
changes only slightly by changing $\Lambda_{R}$, the position of the pole
changes sizably. The precise determination of the pole is not possible at
present, since the value of $\Lambda_{R}$ (as well as the precise form of the
form factor) is unknown. Nevertheless, the existence of a pole in the complex
$\sqrt{s}$ plane is a definite result of our analysis.

\subsection{Dibaryon and $f_{0}(500)$}

In the next step, we add the contribution from $\chi\equiv f_{0}(500)$ [last
line of Eq.\ (\ref{eqLNN})] to that of the dibaryon state of Eq.\ (\ref{eqLR}%
). Using $m_{\chi}=0.525$ GeV and $a=8.95$, it is indeed possible to obtain a
remarkably good agreement with  { SAID results} up to a momentum
$p$ of about $0.4$ GeV, see Fig.\ 5. This shows the importance of the lightest
scalar state $f_{0}(500)$. We recall that this meson is not (predominantly) a
quark-antiquark state [the chiral partner of the pion is identified with the
heavier state $f_{0}(1370)$ \cite{dick}]. Also the description of the
differential cross section is improved, since now the qualitative form is
correctly described (for $p=0.2$ GeV the agreement is also quantitatively
quite good).

\begin{figure}[ptb]
\centering
\includegraphics[width=0.47\linewidth]{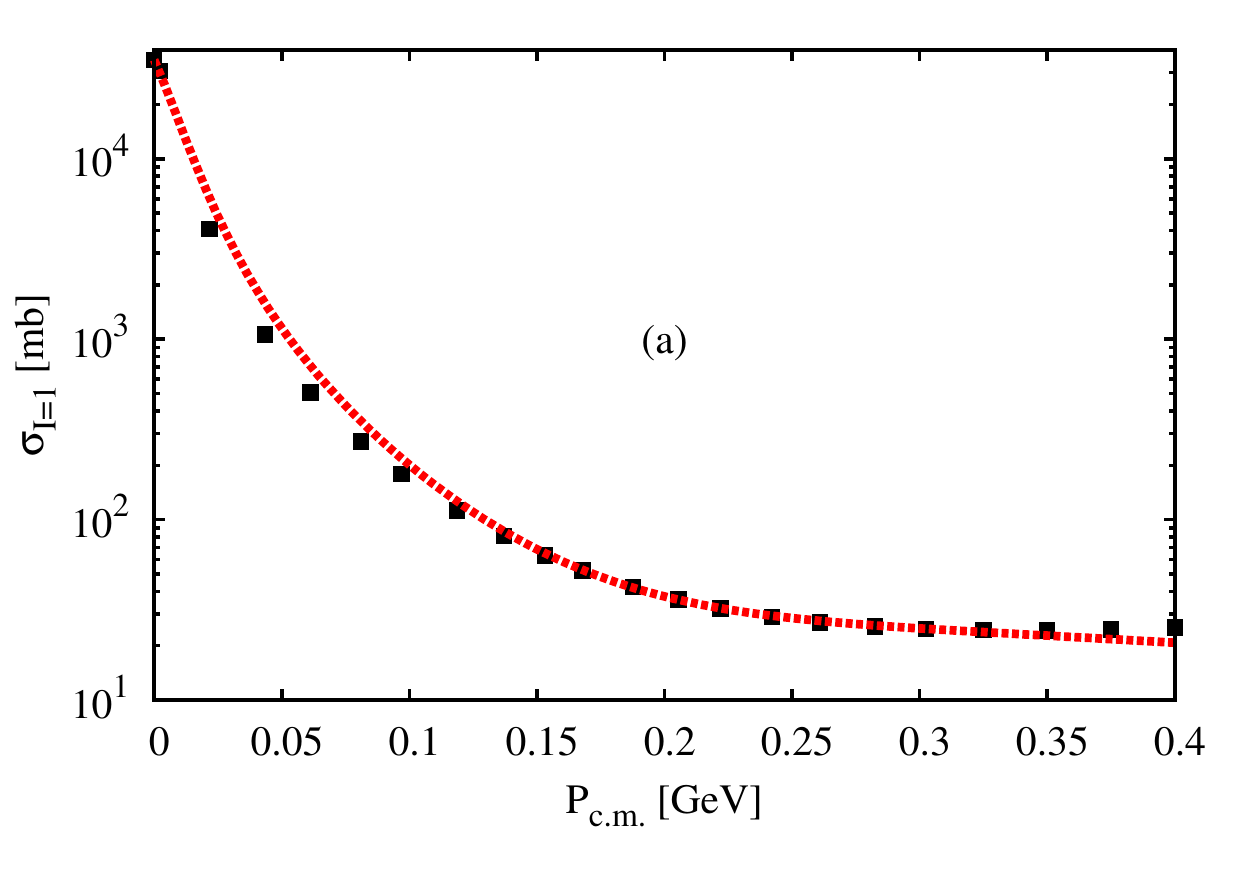} \quad
\includegraphics[width=0.47\linewidth]{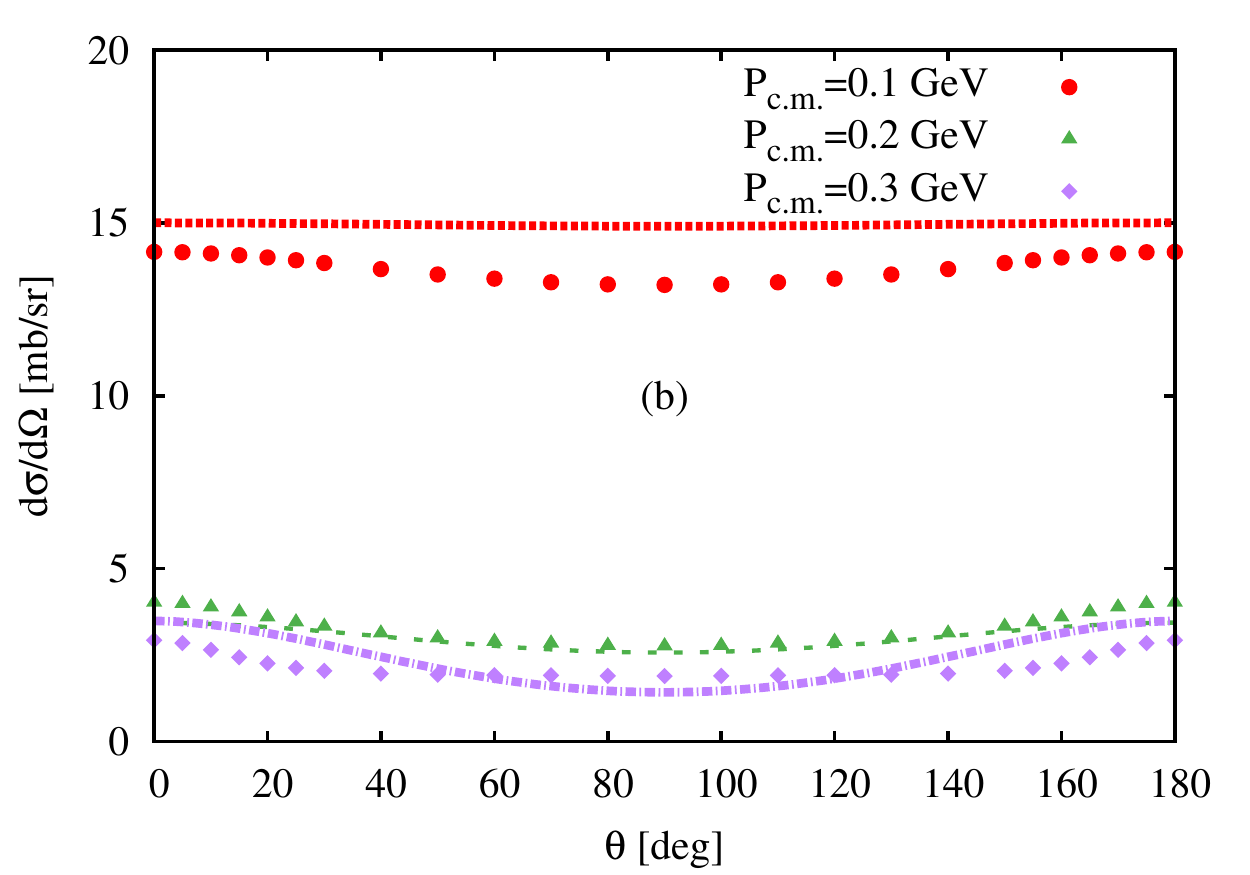}\caption{Total (a)
and differential (b) cross section for $I=1$ $np$ scattering. The
theoretical curve in Fig.\ a is calculated for scattering including
$\chi$ exchange in addition to the contribution of the $^{1}S_{0}$ resonance
with $D_{R}=0.0018~$GeV and $G_{R}=2.27$. The mass of the $\chi$ meson is set
to $0.525$ GeV and its coupling $a =8.95$. In Fig.\ b, the
differential cross section is shown for nucleon c.m.\ momenta 0.1 GeV (red
dotted curve), 0.2 GeV (green dashed curve), and 0.3 GeV (magenta dash-dotted
curve), respectively. Data points are taken from the SAID program
\cite{SAID}.}%
\end{figure}

It is interesting to notice that good agreement with
 { SAID results} is reached for a mass of $f_{0}(500)$ of about
$0.5 - 0.55$ GeV, which is in good agreement with the PDG value. Increasing or
decreasing the value of the mass by about $0.1$ GeV or more considerably
worsens the description of the experimental results.

\subsection{Full model without form factor}

We now turn to the case in which the sum of the two Lagrangians (\ref{eqLNN})
and (\ref{eqLR}) is considered, i.e., when all other mesons are also present.
First, we do not include any form factor in the calculation (i.e.,
$\Lambda_{cut}\rightarrow\infty$). The mass of $f_{0}(500)$ is chosen to be
$m_{\chi}=0.475$~GeV and $a=$ $8.95$.

\begin{figure}[ptb]
\centering
\includegraphics[width=0.47\linewidth]{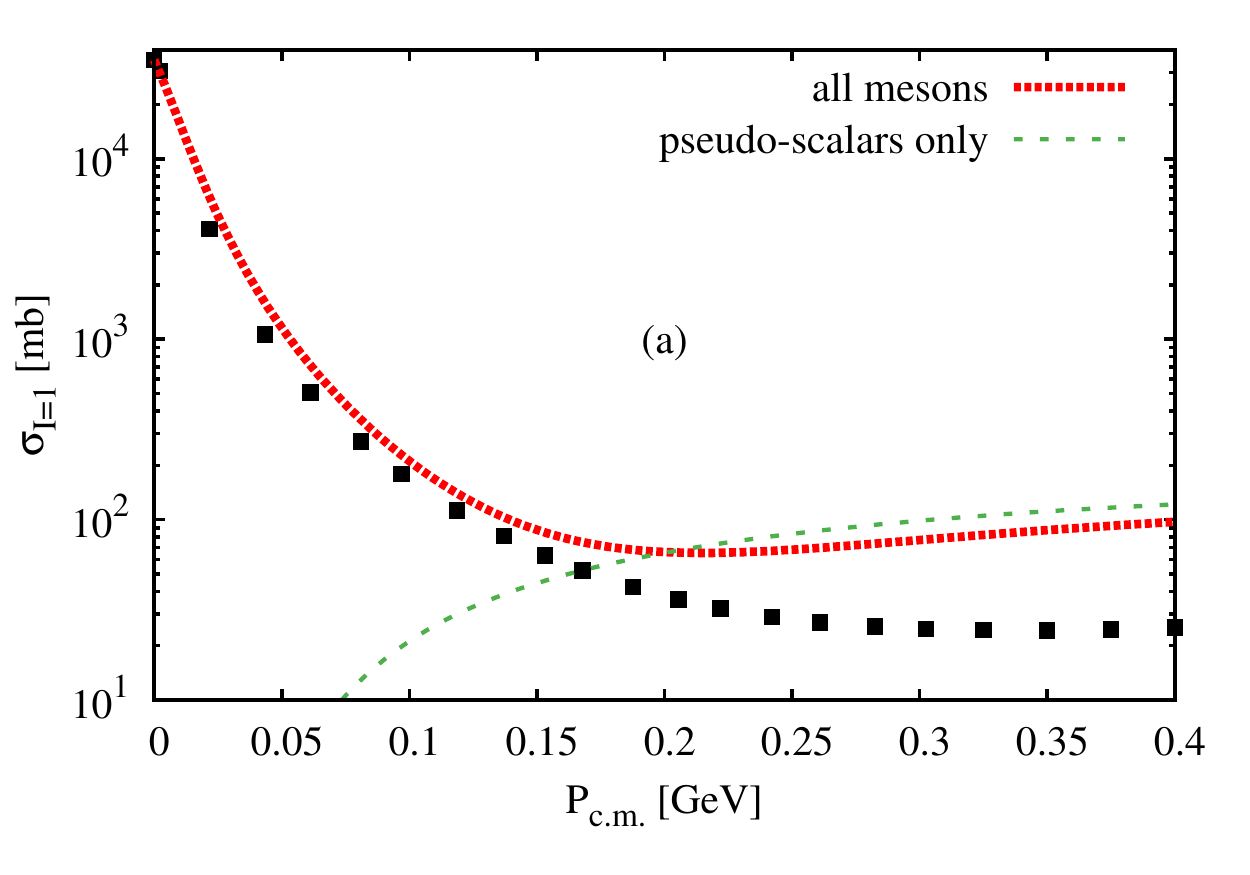} \quad
\includegraphics[width=0.47\linewidth]{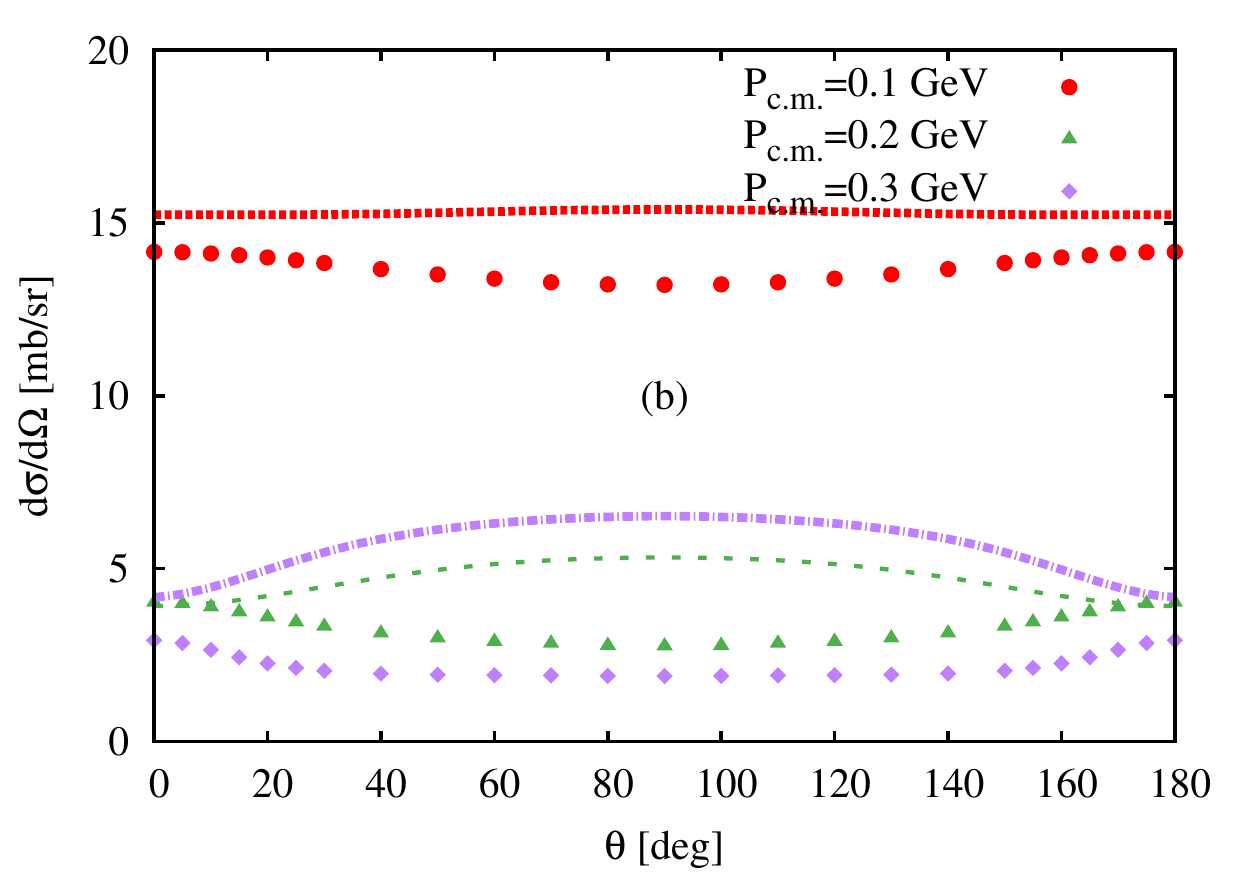}\caption{Total (a)
and differential (b) cross section for $I=1$ $np$ scattering. The
red dotted curve in Fig.\ a is for scattering via exchange of the nine
mesons included in Eq.\ (\ref{eqLNN}) ($m_{\chi}$=0.475~GeV and $a$=8.95) in
addition to the $^{1}S_{0}$ resonance with $D_{R}$=0.0018~GeV and $G_{R}%
$=2.26. The green dashed curve shows the cross section calculated using only
$\pi$ and $\eta$ exchange. In Fig.\ b, the differential cross section
is shown for nucleon c.m.\ momenta 0.1 GeV (red dotted curve), 0.2 GeV (green
dashed curve), and 0.3 GeV (magenta dash-dotted curve) upon exchange of the
nine mesons. Data points are taken from the SAID program \cite{SAID}.}%
\label{fig:res:allnoff}%
\end{figure}

The results are shown in Fig.\ \ref{fig:res:allnoff}. In part (a), one
observes that the theoretically calculated total cross section significantly
overestimates the  { SAID results} for large momenta. The reason
for this is the contribution from the pseudoscalar mesons (green dashed
curve). In addition, the theoretically calculated differential cross sections,
part (b), show the wrong behavior as a function of scattering angle: they are
enhanced at 90$^{0}$, while the { SAID results}
are suppressed.

Apparently, adding the contributions from exchange of quark-antiquark mesons
worsens the agreement with  { SAID results} as
compared to the previous case. One possibility to ameliorate the situation
could be to modify the parameters of the Lagrangian (\ref{eqL1}). Quite
curiously, for $c_{1}=1.5$ the contribution of the pions turns out to be
suppressed due to destructive interference. However, by doing so, one would
inevitably induce a disagreement with other quantities, such as the nucleon
masses and the axial coupling constants. Another possibility, explored in the
following subsection, is to use a finite cutoff, which effectively takes into
account that hadrons are extended objects, and which suppresses the
contribution from pseudoscalar mesons to an extent that the good description
obtained with the ${}^{1}S_{0}$ resonance and $f_{0}(500)$ exchange alone is re-obtained.

\subsection{Full model with form factors}

As a last step we consider both Lagrangians (\ref{eqLNN}) and (\ref{eqLR}) as
well as the form factor introduced in Eq.\ (\ref{ff}). As Fig.\ 7 shows, a
cutoff $\Lambda_{cut}=0.778$ GeV suppresses the contributions of the
quark-antiquark mesons (and in particular pseudoscalar mesons) at large
momenta. The  { SAID results} can be again well described.

The differential cross sections point also to an interesting fact: if the
contribution from $f_{0}(500)$ is turned off, the angular distribution is
again enhanced at 90$^{0}$, in contradiction with
 { SAID results} which are suppressed at this
angle. When $f_{0}(500)$ is taken into account, the correct shape of the
differential cross section is obtained. Apparently, suppressing the
contribution of quark-antiquark mesons by a form factor is not sufficient to
produce the correct angular dependence of the differential cross section, one
needs to include the $f_{0}(500)$ in order to repair this shortcoming. This
confirms once more the important role of this meson for a good description of
nucleon-nucleon scattering.

\begin{figure}[ptb]
\centering
\includegraphics[width=0.5\linewidth]{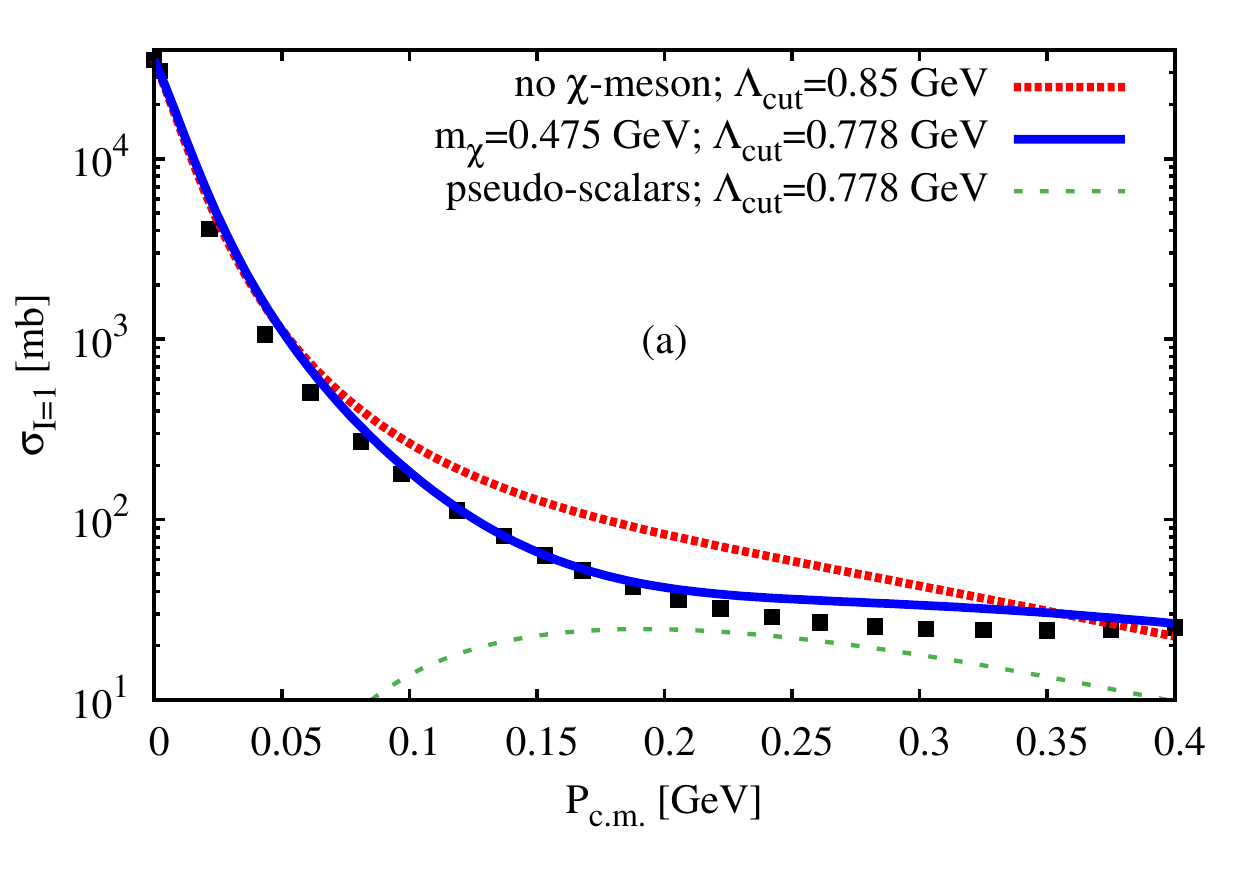} \newline%
\includegraphics[width=0.47\linewidth]{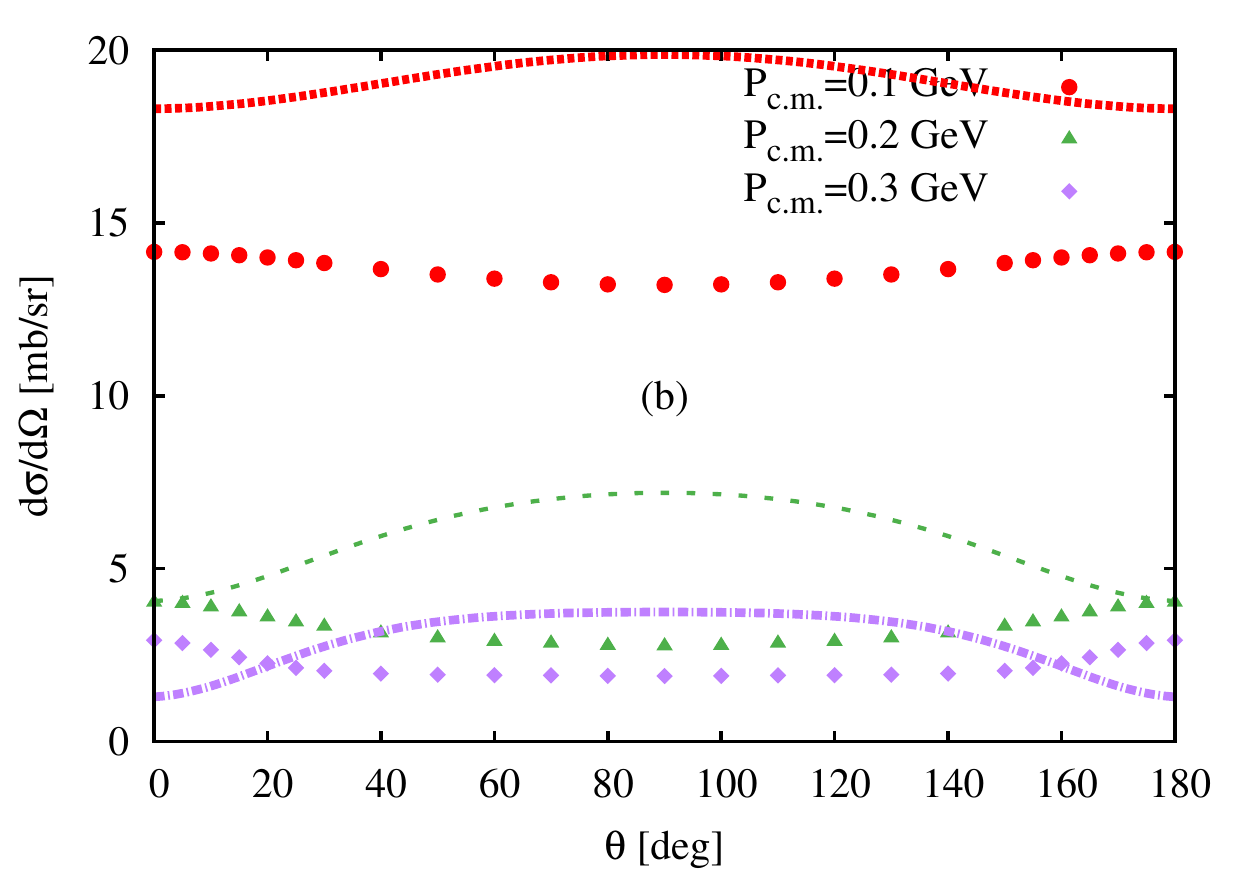} \quad
\includegraphics[width=0.47\linewidth]{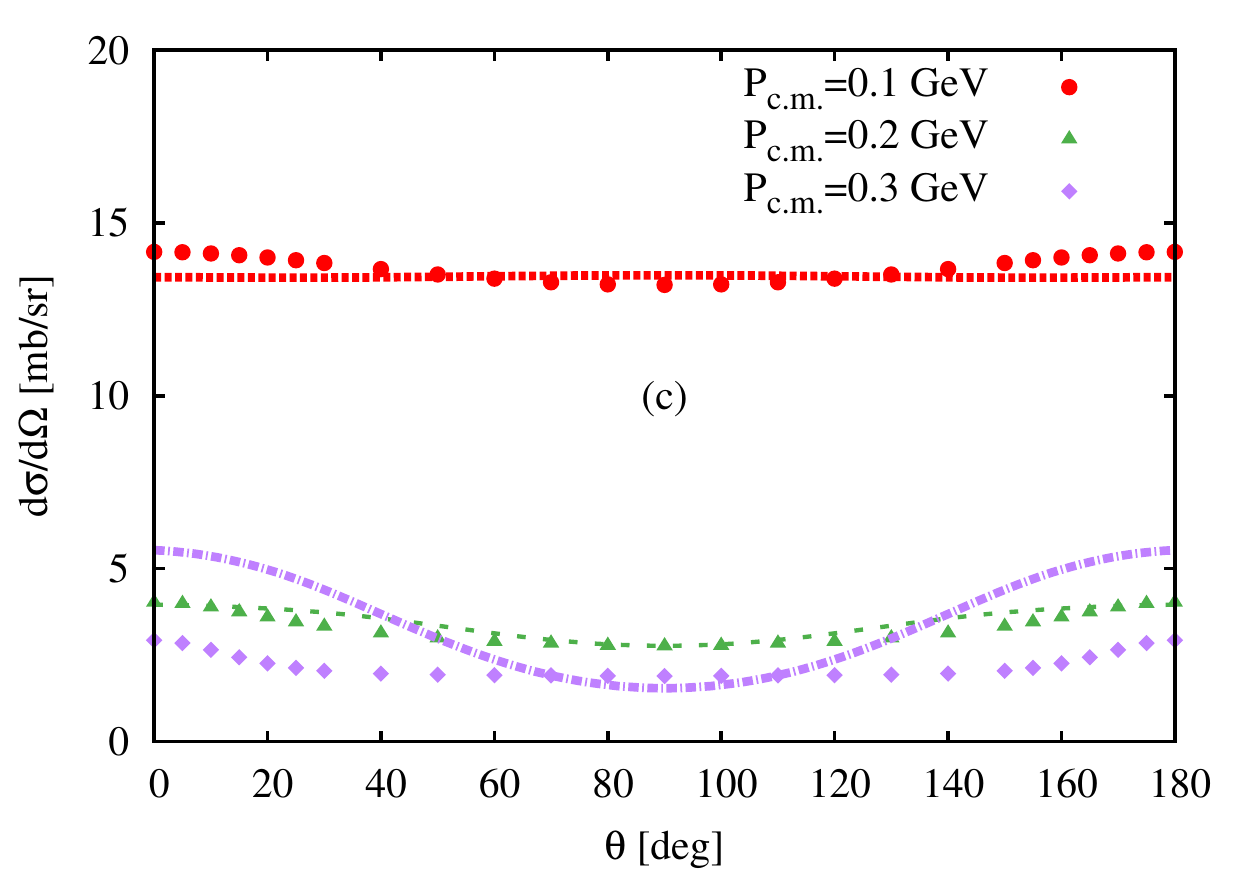}\caption{Total (a)
and differential cross sections (b,c) for $I=1$ $np$
scattering. The red dotted and blue solid curves in Fig.\ a are
calculated for $\sigma,a_{0},\pi,\eta,\omega,\rho,f_{1},a_{1}$ exchange as
well as including the $^{1}S_{0}$ resonance with $D_{R}$=0.0016~GeV. The red
dotted curve shows the case without the $\chi$ meson ($\Lambda_{cut}%
$=0.85~GeV), while the blue solid curve shows the case where a $\chi$ meson
with mass $m_{\chi}$=0.475 GeV and coupling $a$=12.41 is included
($\Lambda_{cut}$=0.778~GeV). The green dashed curve shows the cross section
calculated using only $\pi$ and $\eta$ exchange. Fig.\ a (c)
shows the differential cross section without (with) $\chi$ meson exchange and
$\Lambda_{cut} = 0.85$ GeV (0.778 GeV) upon inclusion of the eight other
mesons, for nucleon c.m.\ momenta 0.1 GeV (red dotted curve), 0.2 GeV (green
dashed curve), and 0.3 GeV (magenta dash-dotted curve), respectively. Data
points are taken from the SAID program \cite{SAID}.}%
\end{figure}

\section{Conclusions}

In this work, we have studied neutron-proton scattering in the $I=1$ channel
in the framework of a chiral model which contains quark-antiquark
(pseudo)scalar and (axial--)vector mesons as well as a scalar isoscalar state
corresponding to the resonance $f_{0}(500)$ (see Fig.\ 1 for the corresponding
diagrams). The $f_{0}(500)$ state is coupled to nucleons in a chirally
invariant way using the mirror assignment for the chiral partner of the nucleon.

The exchange of mesons (Fig.\ 1) alone is not sufficient to describe the very
large total cross section close to threshold. For this reason, we have coupled
the nucleons to a resonance with baryon number $B=2,$ isospin $I=1$, total
spin zero, and positive parity, $J^{P}=0^{+}$. Then, for a suitably chosen
coupling to the nucleons, $s$-channel scattering through this resonance is
able to reproduce the magnitude of the total cross section close to threshold,
see Fig.\ 2.

The results have been presented by including the ingredients step by step. The
total cross section close to threshold can be well described with the help of
the dibaryon alone (Fig.\ 3). A more detailed study of this resonance shows
that it is not of a standard Breit-Wigner type, because the width is larger
than the distance of its mass to the neutron-proton threshold. Its spectral
function (Fig.\ 4) shows a peak very close to threshold. In the complex plane,
we find a pole. For a cutoff of $0.5$ GeV the pole lies at $m_{p}%
+m_{n}+0.014\text{ GeV}-i\; 0.0774\text{ GeV}$, confirming that the resonance
is very broad. However, the pole is not precisely determined because a slight
modification of the parameters changes its position quite substantially.
Nevertheless, the important point is that a pole is always present, which
shows that a dibaryon resonance exists. Interestingly, a similar conclusion
concerning a metastable neutron-proton state was also obtained in
Ref.\ \cite{ivanov} and recent experimental activity is described in
Ref.\ \cite{borzakov}.

As a consequence of our results, we also predict the existence of a
neutron-neutron resonance very close to threshold: this state is the
$I_{z}=-1$ member of the $I=1$ multiplet of scalar dibaryons. The
neutron-neutron resonance is not affected by Coulomb repulsion, thus the
characteristics of the corresponding resonance are expected to be similar to
the proton-neutron dibaryon studied in this work. Indeed, in
Ref.\ \cite{spyrou} a scalar neutron-neutron resonance has been observed
experimentally. The corresponding decay width of about $0.01$ GeV is actually
in good agreement with our results (for the width of the $np$ state, which
should be very similar to the one of the $nn$ state). The subsequent
theoretical study of Ref.\ \cite{hammer} by means of an effective Lagrangian
confirmed that such a dineutron state cannot be excluded. Quite interestingly,
the existence of scalar isotriplet dibaryon may also be relevant in the
context of nuclear astrophysics \cite{macdonald}. Also, the recent discovery
of a four-neutron quasi-bound state \cite{kisamori} shows that the formation
of metastable states made solely of neutrons is possible.

The last member of the isotriplet dibaryon multiplet has $I_{z}=1$ and
consists of two protons. In this channel predictions are more difficult in
view of the Coulomb repulsion that breaks isospin symmetry. However, also here
a resonance could exist, but would be even more unstable, see the experimental
study in Refs.\ \cite{gomezdelcampo,dymov} and theoretical discussion in
Ref.\ \cite{haidennew}.

{In conclusion, in the present work we have found that a pole on the
second Riemann sheet in the $S$-wave }$I=1${ channel is present. As
discussed in Refs.\ \cite{machleidt,vankolck,coraggio,machleidtrev} one does
not need to include an explicit d.o.f.\ in the Lagrangian, since a quartic
interaction together with its resummation would mimic the effect of a
propagator in the }$S${-wave. Within this context, it would be very
interesting if the position of the pole could be also investigated in the
context of such effective approaches. }

Turning back to neutron-proton scattering studied in this work, the next step
has been the inclusion of the resonance $f_{0}(500)$: a remarkably good
agreement with { SAID results} is obtained when only the
dibaryon and the resonance $f_{0}(500)$ are considered (Fig.\ 5). These
results show that these two resonances are most important for the description
of the  { SAID results}.

Switching on the other mesons causes a disagreement at large momenta, because
the contribution of the pions is too large without introducing a form factor
to suppress large momenta. Moreover, also the differential cross sections
cannot be reproduced (Fig.\ 6). This mismatch can, however, be removed by
including a form factor. One then obtains a good description of
 { SAID results} at high momenta (Fig.\ 7). Also in this case,
the role of $f_{0}(500)$ is important: by switching it off, the shape of the
differential cross section is qualitatively wrong.

As an outlook for future studies, one could use our chiral approach to study
reactions in which mesons are produced, such as $NN\rightarrow NNX$ with
\ $X=\omega,\rho,\ldots$ [see Ref.\ \cite{teilabproc} for a preliminary
investigation]. These reactions are at the center of experimental studies, see
e.g.\ Ref.\ \cite{balestra}, and their investigation is important in hadronic
physics. Also similar reactions involving strangeness are relevant: for that
purpose one would need the full version of the eLSM for $N_{f}=3$, including
baryons [a first step towards this goal has been performed in
Ref.\ \cite{lisa}] {as well as the full nonet of light-scalar mesons
below 1 GeV}. For instance, the reaction $pp\rightarrow ppK$ has received
considerable attention \cite{pptopkaon}. The determination of the
baryon-baryon-meson couplings in the three-flavor case is not only relevant
for hadron vacuum physics but also in the context of neutron-star
investigations \cite{hyperonstar}.

\bigskip

\textbf{Acknowledgments}: The authors would like to thank S.\ Gallas,
A.\ Habersetzer, L.\ Olbrich, D.\ Parganlija, and M.\ Zetenyi for their help
during this work. Special thanks go to S.\ Leupold, J.\ Reinhardt, and
J.\ Schaffner-Bielich for very helpful suggestions and to O.\ Mattelaer for
his support in running and debugging MadGraph. We would also like to thank
H.\ Feldmeier, J.\ Gegelia, J.\ Haidenbauer, and H.-W.\ Hammer for useful
discussions. This work was supported by DFG grant no.\ RI 1181/6-1 .
\FloatBarrier

\end{document}